\documentclass{elsart}
\usepackage{natbib}
\usepackage{epsfig}
\newcommand{\sigann}{\ensuremath{\sigma^*_{a}}}
\newcommand{\sigsca}{\ensuremath{\sigma^*_{s}}}
\newcommand{\tstep}{\ensuremath{\Delta t}}
\newcommand{\tdyn}{\ensuremath{t_{dyn}}}
\newcommand{\lcdm}{$\Lambda$CDM}
\newcommand{\nonint}{S0A0}
\newcommand{\scaonly}{S3A0}
\newcommand{\scaann}{S3A1}
\newcommand{\longann}{S.3A.1}
\newcommand{\colhead}[1]{#1}
\newcommand{\sun}{\odot}
   
\newcommand\apj{{ApJ}}% 
\newcommand\apjl{{ApJ}}% 
\newcommand\mnras{{MNRAS}}% 
\newcommand\prd{{Phys.~Rev.~D}}% 
\newcommand\prl{{Phys.~Rev.~Lett.}}% 
\newcommand\pasj{{PASJ}}% 
\newcommand\nat{{Nature}}% 
\begin{document} 
\begin{frontmatter}
\title{The Structure of Dark Matter Halos in an
  Annihilating Dark Matter Model}
\author{Matthew W. Craig\corauthref{cor}} 
\corauth[cor]{Corresponding author.} 
\address{Department of Physics and Astronomy,
Minnesota State University Moorhead, Moorhead, MN 56563}
\ead{mcraig@mnstate.edu}

\author{Marc Davis}
\address{Astronomy and Physics Departments, University of California,
  Berkeley CA 94720} 
\ead{marc@astro.berkeley.edu}
\begin{abstract}
  The inability of standard non-interacting cold dark matter (CDM) to
  account for the small scale structure of individual galaxies has led
  to the suggestion that the dark matter may undergo elastic and/or
  inelastic scattering. We simulate the evolution of an isolated
  dark matter halo which undergoes both scattering and annihilation.
  Annihilations produce a core that grows with time due to adiabatic
  expansion of the core as the relativistic annihilation products flow
  out of the core, lessening the binding energy. An effective
  annihilation cross section per unit mass equal to
  .03~cm$^2$~g$^{-1}$ (100~km~s$^{-1}$/$v$) with a scattering cross
  section per unit mass of .6~cm~g$^{-1}$ produces a 3~kpc core in a
  10$^{10}$~M$_{\odot}$ halo that persists for 100 dynamical times.
  The same cross section leads to a core of only 120~pc in a rich
  cluster. In addition to creating to cores, annihilation should erase
  structure on scales below $\sim 3\times10^8$~M$_\odot$. Annihilating
  dark matter provides a mechanism for solving some of the problems of
  non-interacting CDM, at the expense of introducing a contrived
  particle physics model.
\end{abstract}

  
\end{frontmatter}

\section{Introduction}
Discrepancies between the Cold Dark Matter (CDM) model and
observations on galactic scales
\citep{salucci:conscore:2001,moore:dwarfs:1994,flores:dwarfs:1994,burkert:dwarfstruct:1995,salucci:dmscale:2000,
  moore:dmcrisis:2001,sellwood:dmexist:2000,weiner:n4123a:2001}
have led  to  recent revival of the suggestion that the dark matter is
self-interacting \citep{spergel:sidm:2000}.

Simulations of halo formation in a CDM universe predict that dark
matter halos have a cuspy inner profile. Most simulations find that,
at small radii, $\rho \propto r^\alpha$, with $\alpha \approx -1$ to
$-1.5$ \citep{navarro:halostruct:1996, 
  ghigna:innherhalo:2000,jing:halo-inner:2000,
  klypin:inner-halo:2000}.  An inner density profile this steep may be
consistent with the dark matter density profile inferred from many
high surface brightness galaxies and clusters of galaxies
(\cite{eke:new-halo-den:2000}, but see \cite{borriello:dmdisc:2001} ).
However, low surface brightness galaxies, and dwarf irregular galaxies
in particular, require an inner density profile which is less steep
(\cite{moore:dwarfs:1994, burkert:dwarfstruct:1995,
  flores:dwarfs:1994}; for a contradictory view see
\cite{bosch:dwarf-cores:2000}).  In addition, \citet{debattista:bars-core:2000}
have emphasized that too much dark matter in the centers of galaxies
is incompatible with rotating bars, which are common in massive
galaxies such as the Milky Way.

A separate problem for standard CDM model is that it predicts many
more dark matter satellites of a Milky Way-sized galaxy than are
observed \citep{moore:substruct:1999,klypin:substruct:1999}. This
substructure cannot simply be dark satellites in which stars have not
formed, because they would thicken the disk of the Milky Way as they
passed through the plane of the disk.

While it is possible that the inclusion of baryonic physics would lead
to a less cuspy dark matter halo, perhaps due to supernova driven gas
outflow in dwarf galaxies \citep{navarro:dwarfcore:1996}, this seems
unlikely, as cooling baryons would lead to further concentration of
the dark matter halo through adiabatic compression. Exploding gaseous
disks will drive a strong wind off the plane, but most simulations
show that the cold, dense component of a multi-phase medium would be
difficult to remove \citep{maclow:winds:2001}. Furthermore, these
solutions to the cusp problem does not address the problem of too much
substructure on small scales.

Recent attention has focused instead on the possibility that the dark
matter is self-interacting. A remarkable number of different models
have been proposed. For example, several authors have considered a
model in which the dark matter is a self-interacting scalar field with
quartic potential \citep{peebles:fluiddm:2000, peebles:fluiddm:1999,
  goodman:fluiddm:2000, riotto:quartic:2000}. \citet{hu:fuzzydm:2000}
have suggested a model in which cores are produced in dwarf galaxies
due to the wave nature of dark matter, if the matter is an ultra-light
scalar particle.  Another interacting dark matter candidate is
Q-balls, non-topological solitons with properties that allow it to
scatter, possibly with a time-dependent cross-section, or to merge
\citep{kusenko:q-ball:2001}.  Several authors have simulated halo
evolution and formation in scenarios in which the dark matter
scatters. All authors agree that on short time scales, scattering
reduces the logarithmic slope of the inner density profile
substantially; some authors have found that the inner density profile
subsequently steepens as the halo undergoes core collapse
\citep{kochanek:interact:2000,yoshida:scatter:2000}, while others disagree
\citep{burkert:interact:2000,dave:sidm:2001}.

Self-annihilating dark matter is another possible solution to the
problem \citep{kaplinghat:2000}, although the particle physics must be
carefully contrived so that the dark matter does not completely
self-annihilate in the early universe. An analytic calculation of the
density of a self-annihilating dark matter halo demonstrates the
formation of a core. Physically, the core develops because
annihilation becomes very efficient at removing mass from the halo
above a density that depends on the annihilation cross section.
\citet{kaplinghat:2000} point out that this annihilation also causes
an adiabatic expansion of the halo, and numerically calculate the
effects of adiabatic expansion; the core density including expansion
is smaller than that predicted from annihilation alone by a factor of
ten (for dwarf galaxies) to three (for clusters).  In addition,
\citet{hui:unitarity:2001} points out that unitarity bounds require
that annihilation be accompanied by scattering.

In this paper, we simulate the behavior of an isolated dark matter
halo which undergoes both scattering and annihilation. In section
\ref{sec:DMModel}, we describe our dark matter model and our
simulation methods. In section \ref{sec:Results} we present the results of our
simulations, and in section \ref{sec:Conc} we compare our results with
those of previous authors.

\section{Model}
\label{sec:DMModel}

\subsection{Initial halo and units}
We simulated the evolution of an isolated halo; the initial density
profile is the \cite{hernquist:profile:1991} profile,
\begin{equation}
  \label{eq:hern-prof}
  \rho = \frac{M}{2\pi r (r+r_s)^3},
\end{equation}
where $M$ is the total mass of the halo, and $r_s$ is the scale
radius.  We take the dynamical time for this model to be the circular orbit
time at the scale radius $r_s$,
\begin{equation}
  \label{eq:tdyn-def}
  \tdyn = 4\pi \sqrt{\frac{r_s^3}{GM}}
\end{equation}
In the simulations below, we work in units in which $G=1$, $M=1$, and
$r_s=1$. 

The profile has been generated to guarantee that the center of mass of
the halo is also the origin used in calculating the density of the
Hernquist profile. We first choose the energy for each particle, then
select a radius for the particle; these determine the magnitude of the
velocity of the particle. We adopt an isotropic velocity distribution.
For each particle generated with position $\mathbf{r}$, a second
particle is generated with position $-\mathbf{r}$; the second particle
has the same energy as the first.  The distribution is truncated at a
radius of 100$r_s$.  The direction of the velocity of each particle is
generated independently, but the net velocity of the halo is set to
zero.

\subsection{Model for scattering and annihilation.}

We adopt a model similar to that used by \citet{burkert:interact:2000}. The probability for interaction for a
particle which moves a distance $\Delta x$ in time step $\tstep$ is
\begin{equation}
  \label{eq:prob-def}
  P = \frac{\Delta x}{\lambda},
\end{equation}
where $\lambda=1/(n\sigma)$ is the mean free path for interactions;
this expression is valid only when $\Delta x \ll \lambda$.  Our
simulations include both scattering and annihilation. We use the
notation $\sigann$ to denote the annihilation cross section per unit
mass and $\sigsca$ to denote the scattering cross section per unit
mass.

The probability that a \emph{pair} of particles scatters in time step
$\Delta t$ is 
\begin{equation}
  \label{eq:scat-prob}
  P_{s,ij} = \sigsca v_{ij} \rho_l \Delta t,
\end{equation}
where $\rho_{l}$ is the local density, and $v_{ij}$ is the
\emph{relative} velocity of particles $i$ and $j$. Scattering is
completely elastic, and is isotropic in the center of momentum frame
of the pair of particles.

As pointed out by \cite{kaplinghat:2000}, only $s$-wave annihilation is
consistent with observations. For $s$-wave annihilation, $\sigma |v|$
is independent of velocity, so that the cross section for annihilation
per unit mass, $\sigma^*_a$, is given by
\begin{equation}
  \label{eq:ann-xsec-def}
  \sigma_a^* = \frac{\tilde{\sigma}_a}{m} \frac{v_a}{v},
\end{equation}
where $\tilde{\sigma}_a {v_a}$ is determined by the scattering
potential. The numerical value we assume for the scattering cross
section is given below.

The probability for an annihilation interaction in time step
$\Delta t$ is:
\begin{equation}
  \label{eq:prob-ann}
  P_{a} = \sigann \rho_{l}  v \Delta t 
         = \frac{\tilde{\sigma}_a v_a}{m} \rho_{l}  \Delta t,
\end{equation}
where $\rho_{l}$ is the local density. The annihilation
probability is independent of velocity. 

We model the decay of the dark
matter particles represented by our simulation particles by assigning
to each simulation particle a probability for annihilation given by
equation (\ref{eq:prob-ann}). If the local density is large enough,
$P_{a}$ is the probability that some of the dark matter particles
represented by the simulation particle would interact and decay.  If a
simulation particle decays, its mass $m_i$ is reduced by fraction $F$,
\begin{equation}
  \label{eq:mass-change}
  m_{i, new} = (1-F) m_{i, old}.
\end{equation}
Thus, the effective probability for annihilation
of individual dark matter particles in our model is the product
$FP_a$. The mass lost in an annihilation is a pre-set fraction of
the current mass of the particle undergoing annihilation. In other
words, the \emph{fractional} mass lost in an annihilation is held
fixed. The absolute mass loss, $m_{i,old} - m_{i,new}$, decreases 
as a particle undergoing repeated annihilation loses mass.

\subsection{Numerical Parameters}
Our simulations were run on a GRAPE3-AF special purpose hardware card
\citep{okumura:grape3:1993} using a direct summation code. The time
step was chosen to be small enough to satisfy both the Courant
condition and to ensure that $v\Delta t < \lambda$.  Each run
contained 30,000 particles. Detailed numerical parameters for each run
are shown in Table \ref{table:num-params}. Runs \nonint, \scaonly, and
\scaann\ ran for roughly 10
dynamical times with approximately 2100 steps per dynamical time; a
fourth run, \longann, was done with a much larger Plummer softening and time
step; it ran for 100 dynamical times with approximately 400 steps
per dynamical time.

\begin{table}
\caption{Simulation Parameters\label{table:num-params}}
\begin{tabular}{lllllll}
%\tablewidth{0pt}
%\tablehead{%
\colhead{Name} & \colhead{$\sigsca$} & \colhead{$\sigann v$} 
& \colhead{$\epsilon$} & \colhead{$\Delta t$}  & \colhead{$N_{steps}$}
& \colhead{$t_f/t_{dyn}$} \\ \hline
\nonint & 0 & 0 & 0.01 & 0.006 & 20,000 & 9.55\\
\scaonly & 3.0 & 0 & 0.01 & 0.006 & 15,750 & 7.52\\
\scaann & 3.0 & 1.5 & 0.01 & 0.006 & 20,000 & 9.55\\
\longann & 0.3 & 0.15 & 0.05 & 0.03 & 40000& 95.5\\
%\tablecomments{See the text for a discussion of units. All
%  annihilation runs had $F=0.05$.}
\end{tabular}
\end{table}

One run, \nonint, had no interactions, and was used to determine the
stability of the integration. The run \scaonly\ had no annihilations,
but non-zero scattering cross-section. Two runs with annihilation and
scattering 
were performed. One, \scaann, was short (the final time was a tenth of a Hubble
time assuming a 10$^{10}$M$_{\odot}$ halo), but had the same
scattering cross-section as the scattering-only run. The other,
\longann, ran for ten times longer, with both annihilation and
scattering cross section reduced by a factor of ten.  In both
annihilation runs the scattering and annihilation \emph{probabilities}
are equal at the scale radius; inside of that radius
scattering will be enhanced relative to annihilation because the
scattering probability increases with velocity. The fractional mass
loss per annihilation in both was $F=5\%$.

The cross section per unit mass for scattering can be calculated in
physical units from
\begin{equation}
  \label{eq:sigma-phys-scat}
  \sigma^*_{phys,s} = .48 \sigma^*_s
  \left(\frac{10^{10}M_{\sun}}{M}\right)
  \left(\frac{R_s}{1 \mathrm{kpc}}\right)^2~\mathrm{cm^2 g^{-1}}.
\end{equation}
%$
The cross-section for annihilation is given by 
\begin{equation}
  \label{eq:sigma-phys-ann}
  \sigma^*_{phys,a} = .98 \sigma^*_a
  \left(\frac{10^{10}M_{\sun}}{M}\right)
  \left(\frac{R_s}{1 \mathrm{kpc}}\right)^2
  \left(\frac{100 \mathrm{km s^{-1}}}{v}\right)~\mathrm{cm^2 g^{-1}}
\end{equation}
For $R_s=2$~kpc and $M=10^{10}$M$_{\sun}$, the half mass dynamical
time is .168~Gyr. Our cross-sections are $\sigsca = 5.76 \mathrm{cm^2 g^{-1}}$
and $\sigann = 5.88 \mathrm{cm^2 g^{-1}} (100\mathrm{km s^{-1}}/v)$ for the run \scaann;
in the run \longann, both cross sections are a factor of ten smaller.
Our values of \sigsca\ span the range suggested by
\citet{wandelt:sidm:2000}. \citet{kochanek:interact:2000} considered a
broader range of values for \sigsca; our choice for our runs with the
largest scattering cross section is in the middle of their range.

In both annihilating runs, we chose our annihilation cross section to
be comparable to the scattering cross section. However, this
annihilation cross section in the run \scaann\ is about 300 times
larger than that suggested by \citet{kaplinghat:2000} based on fits of
rotation curves to dwarf galaxies. The run \longann, with an
annihilation cross section ten times smaller, is closer to what they
suggest.  With a mass loss factor $F$ of 5\%, the effective cross
section $F\sigann$ is comparable to the value required to match
observations. In both cases, the scattering cross section is above the
minimum required by unitarity bounds \citep{hui:unitarity:2001}.

\section{Results}
\label{sec:Results}

\begin{figure}[htbp]
%\epsscale{.8}     
  \epsfig{file=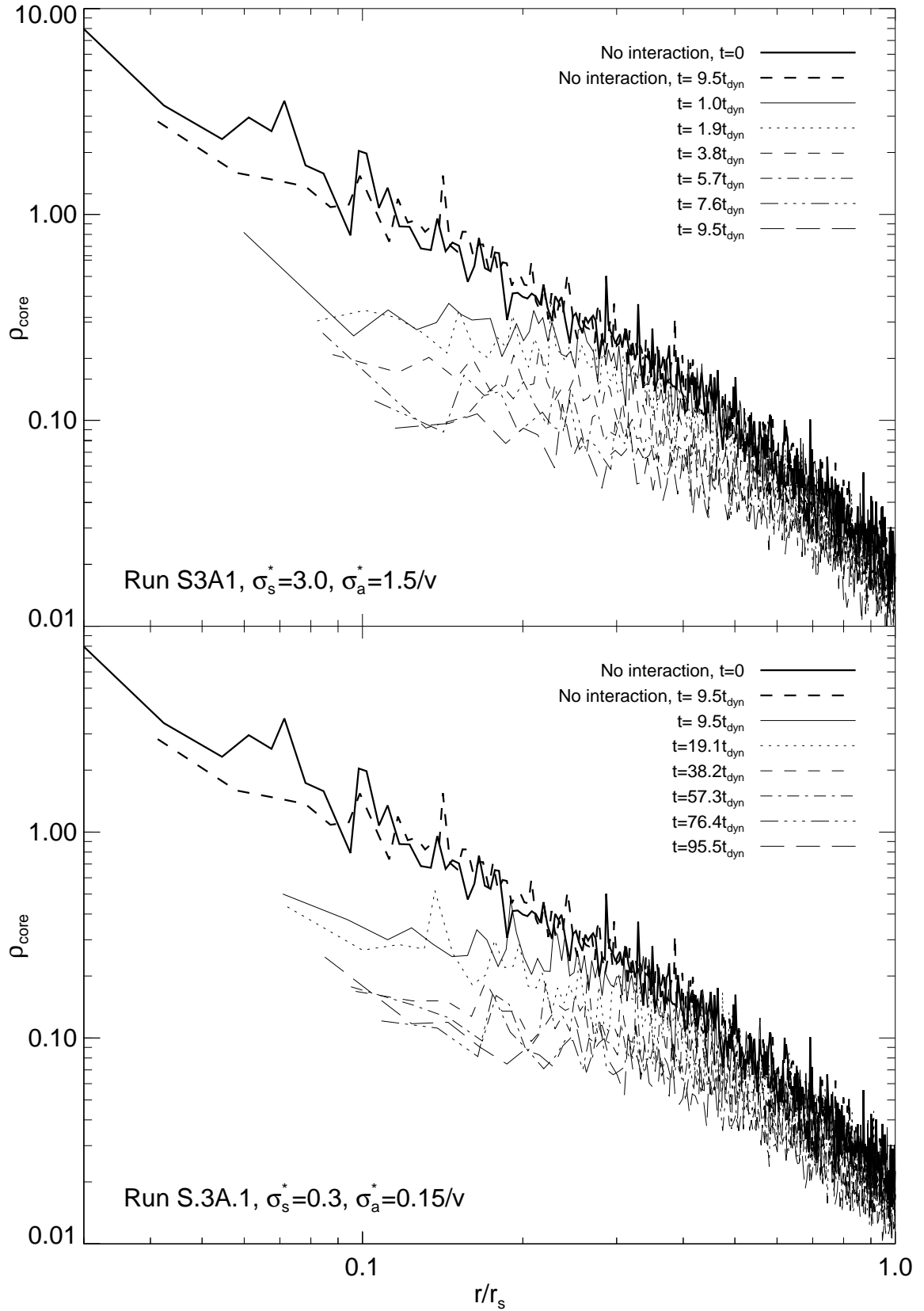}
%  \plotone{fig-ann-den.eps}
  \caption{Evolution of density profile in two models with annihilation and
    scattering. The heavy lines in both plots show the change in core
    behavior without interactions.  In the top plot, the force
    softening length is 0 .01; in the bottom plot it is 0.05. In this
    and later density plots, only the central region of
    the halo is shown; the density profile does not change
    significantly over the course of the simulations for radii larger
    than the scale radius.}
  \label{fig:dens-annihil}
\end{figure}

The inclusion of annihilation in our simulations causes a reduction in
the central density of the halo. Figure \ref{fig:dens-annihil} shows
the evolution in the density profile of a halo with both annihilation
and scattering; the top panel shows a short run with high cross
section (\scaann), and the bottom a run with smaller cross section run
for longer time (\longann). The central density rapidly decreases, and remains low
for the duration of the simulations, about 100 (10) dynamical times
for run \longann\ (\scaann). 

The core is not due to numerical effects. The force softening is .05
(.01) in the run \longann\ (\scaann); the flattening of the inner
density profile we observe occurs at radii 2--10 (10--50) times larger
than that and is unlikely to have been caused by softening.  To test
the stability of our code and the effects of our relatively limited
particle number, we ran a simulation with no interactions for ten
dynamical times. The simulation with no interactions \nonint\ shows no
evidence of the development of a core.  The density profile of the
halo was essentially unchanged over the course of the simulation.
Figure \ref{fig:dens-annihil} shows the beginning profile and the
final profile for this run.
 
We have examined the mechanism by which the core forms in our
interaction simulations. There are three related mechanisms one might
imagine contributing to the development of a core.  First, mass lost
due to annihilation and the disappearance of the relativistic decay
products from the halo directly decreases  the central
density of the halo. Second, this direct mass loss reduces the
gravitational potential of the halo, causing particles in the central
region to become less bound. Third, energy is transferred from the
central region of the halo outward by scattering, depressing the
central density.

\begin{figure}[htbp]
  \epsfig{file=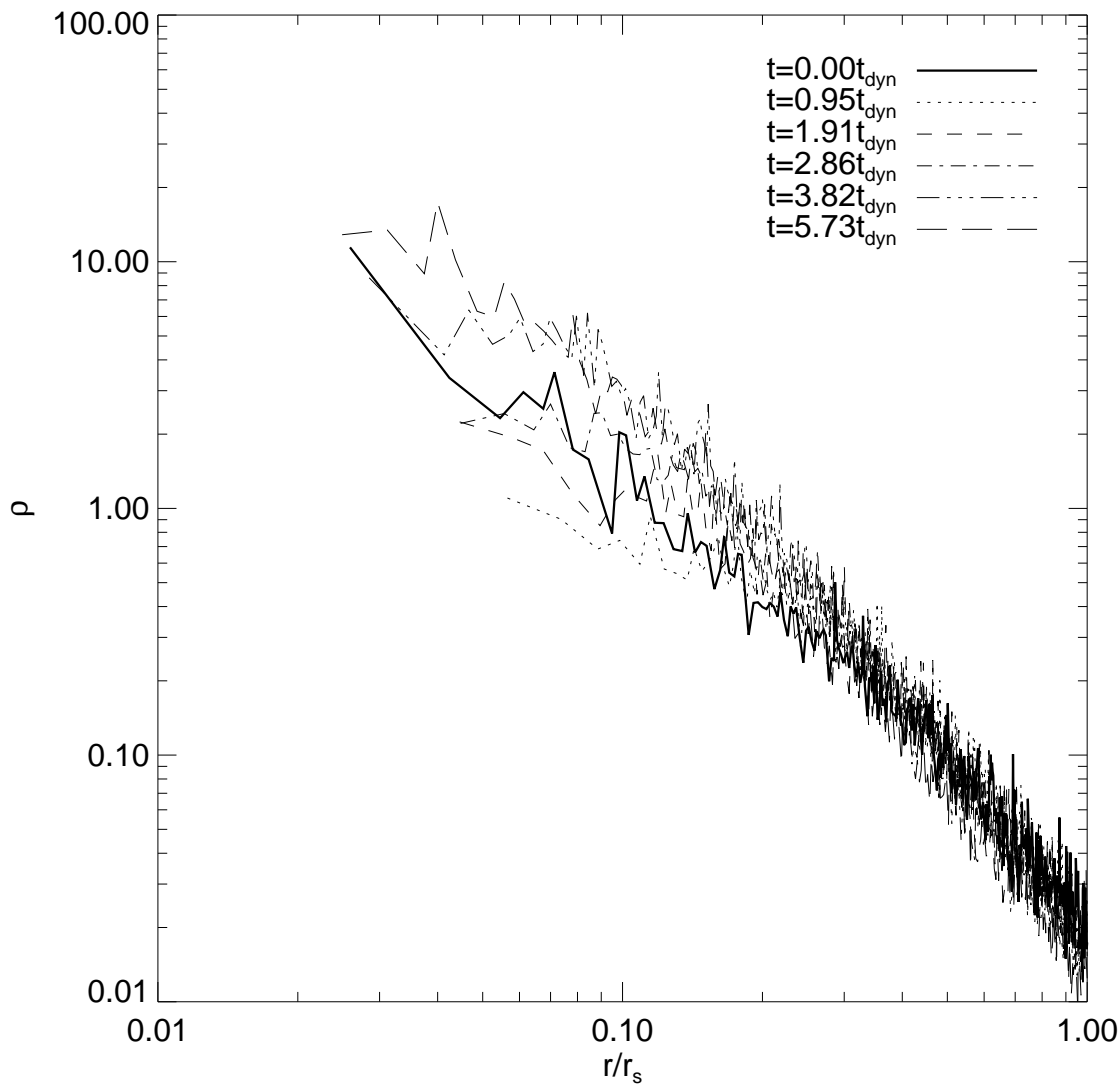}
  \caption{Evolution of density profile in the scattering-only model
    \scaonly. The core density initially decreases, but then begins to
    grow with time.}
  \label{fig:scatt-dens}
\end{figure}

Although scattering at first suppresses the central density, core collapse
begins fairly rapidly in halos with scattering only, as reported by
\cite{kochanek:interact:2000}. We have run one simulation with scattering alone to
determine whether the core that initially forms persists. The results
of this run are shown in Figure \ref{fig:scatt-dens}. The central density initially drops and then increases
after a few dynamical times.\footnote{This run was terminated after
  5.8 dynamical times. At that point, the central density became so
  large that the typical particle was moving more than one mean
  free path in a time step.}
In the absence of annihilation, scattering does not produce a
persistent core in simulations of \emph{isolated} halos. It is unclear
whether or not cores persist in high resolution simulations of halo
formation in a cosmological setting
\citep{yoshida:scatter:2000,dave:sidm:2001}. 

\begin{figure}[htbp]
  
  \epsfig{file=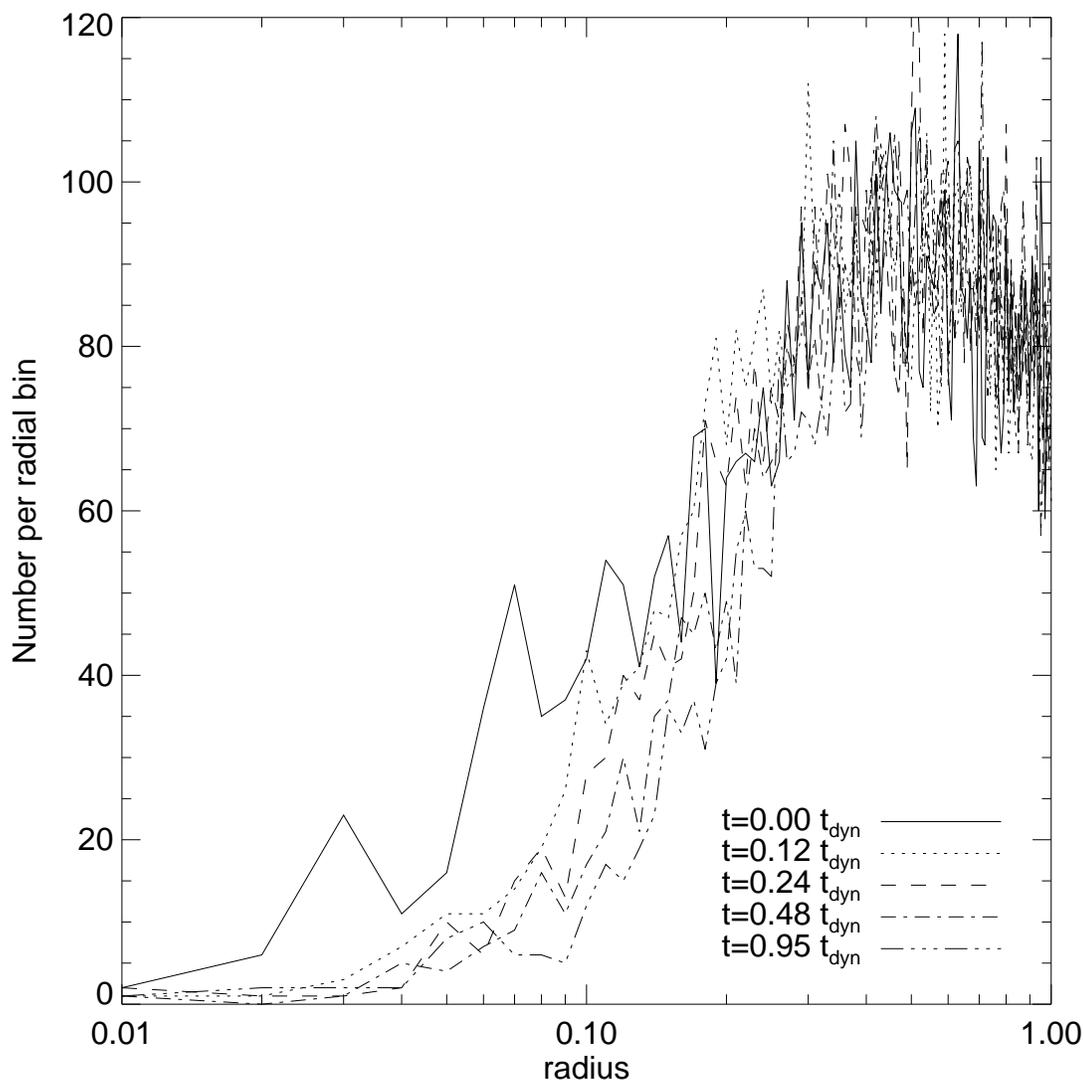}
  \caption{Evolution in number density of particles in annihilation+scattering model.}
  \label{fig:num-dens-change}
\end{figure}

In order to differentiate between direct mass loss from annihilation
and a change in the density profile due to a change in gravitational
potential, we calculate the change in number density of simulation
particles as a function of radius. If direct mass loss due to
annihilation is causing most of the decline seen in the density
profile, the number density profile ought to be unchanged, with
particles in the central region of the halo simply becoming less
massive. Instead, as can be seen in Figure \ref{fig:num-dens-change},
which shows the evolution in the number density of particles early in
the simulation, we find that the number density rapidly diminishes in
the center of the halo, indicating that much of the change in central
density is due to the motion of particles away from the center, not
from direct mass loss. The trend of particle number density decreasing
in the center continues throughout the simulation, as can be seen from
Figure \ref{fig:dens-annihil}. Density profiles are calculated with a
fixed number of particles per bin, so the location of the first point
in the density profile is a radius that encloses a fixed number of
particles; this point moves farther from the origin throughout the
simulation. Furthermore, the total mass lost by the halo over the
entire simulation is only 12\% of the initial halo mass.

\begin{figure}[htbp]
  \epsfig{file=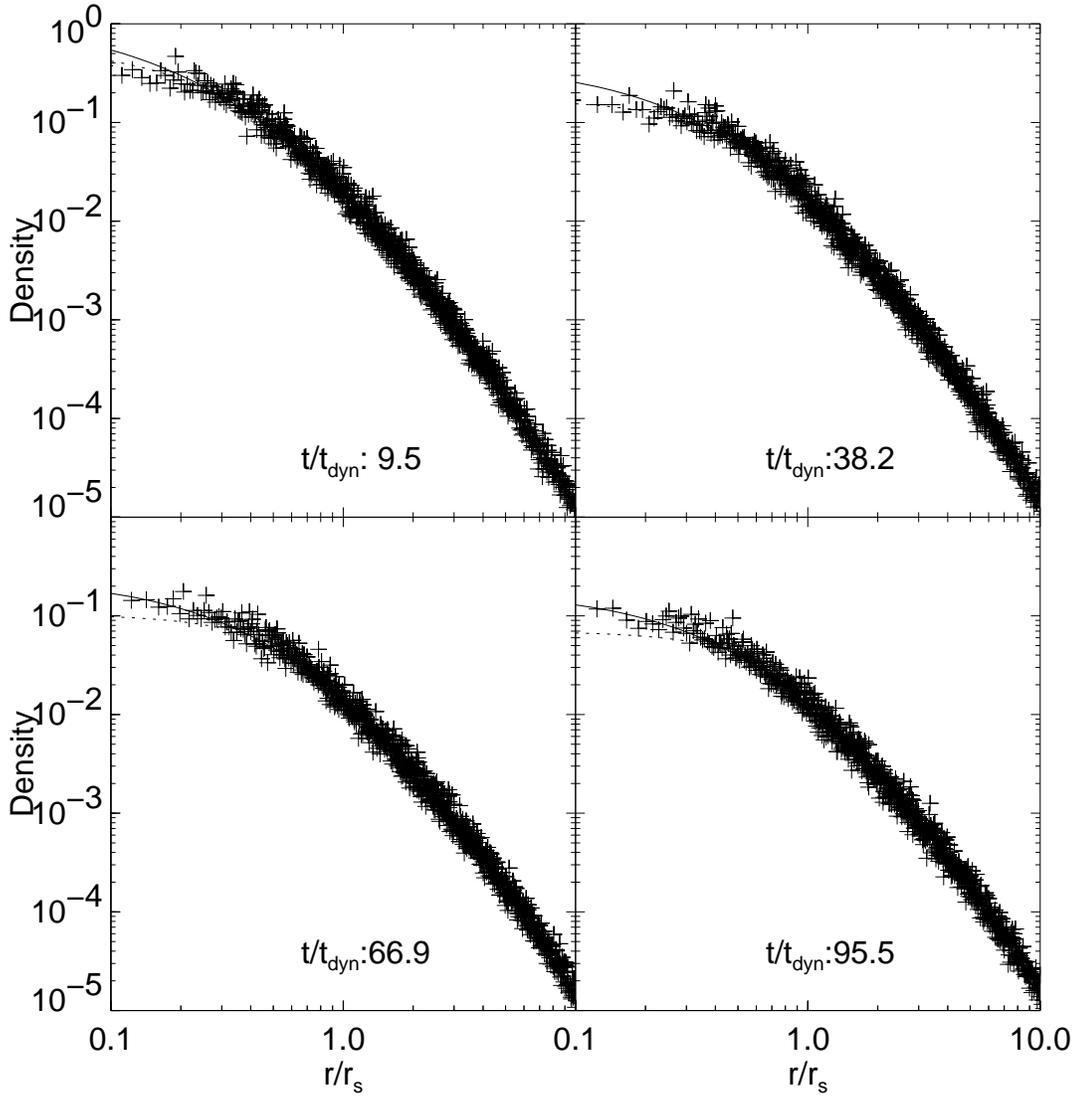}
  \caption{Fits to the density profile of the annihilating dark matter
    run \longann. The solid line shows the best fitting modified
    Hernquist profile, the dotted line the best-fitting Kaplinghat
    profile, and the data are the symbols. As the simulation proceeds,
    the halo is better fit by a modified Hernquist profile than a
    Kaplinghat profile.}
  \label{fig:halo-fits}
\end{figure}

The halo in the annihilating model forms a core very rapidly, and the
size of the core continues to grow over the course of the simulation.
We fit two different functional forms to the density profile of the
halo. One is a modification of the initial Hernquist profile
\ref{eq:hern-prof} to allow for a core,
\begin{equation}
  \label{eq:mod-hern-fit}
  \rho_C = \frac{\rho_0 x_c}{(x+x_c)(1+x)^3},
\end{equation}
where $x=r/r_s$, and $x_c = r_c/r_s$ is the core radius in units of
the scale radius. The other density profile was derived by
\cite{kaplinghat:2000}; their profile takes into account modifications
of the initial halo density profile due to direct annihilation only,
without scattering.  Adiabatic expansion of the core as mass is lost
decreases the core density. We have derived the profile appropriate
for an initial Hernquist model,
\begin{equation}
  \label{eq:kaplin-fit}
  \rho = \frac{\rho_0 x_c}{x(1+x)^3 + x_c}
\end{equation}
(note that $x_c$ is related to the annihilation cross section in
Kaplinghat's model (Eq. \ref{eq:kapl-core-rad}); in
doing our fits we have treated $x_c$ as a free parameter to see how
well the shape of the density profile is fit by the function). Fits to
both functions are shown at four different epochs in Figure
\ref{fig:halo-fits}. The two profiles differ only in the innermost
region of the halo. \citet{burkert:dwarfstruct:1995} found that dwarf
galaxies are well fit by the profile 
\begin{equation}
  \label{eq:burkert-profile}
  \rho_B = \frac{\rho_0 r_c^3}{(r+r_c)(r^2+r_c^2)},
\end{equation}
where $\rho_0$ is the central density of the halo, and $r_c$ its core
radius.  We do not expect the Burkert profile to be a good fit to our
halo because the initial halo profile has both a different radial
dependence at large radii, and a length scale (the scale radius) in
the initial conditions separate from the core radius.

\begin{figure}[htbp]
  \epsfig{file=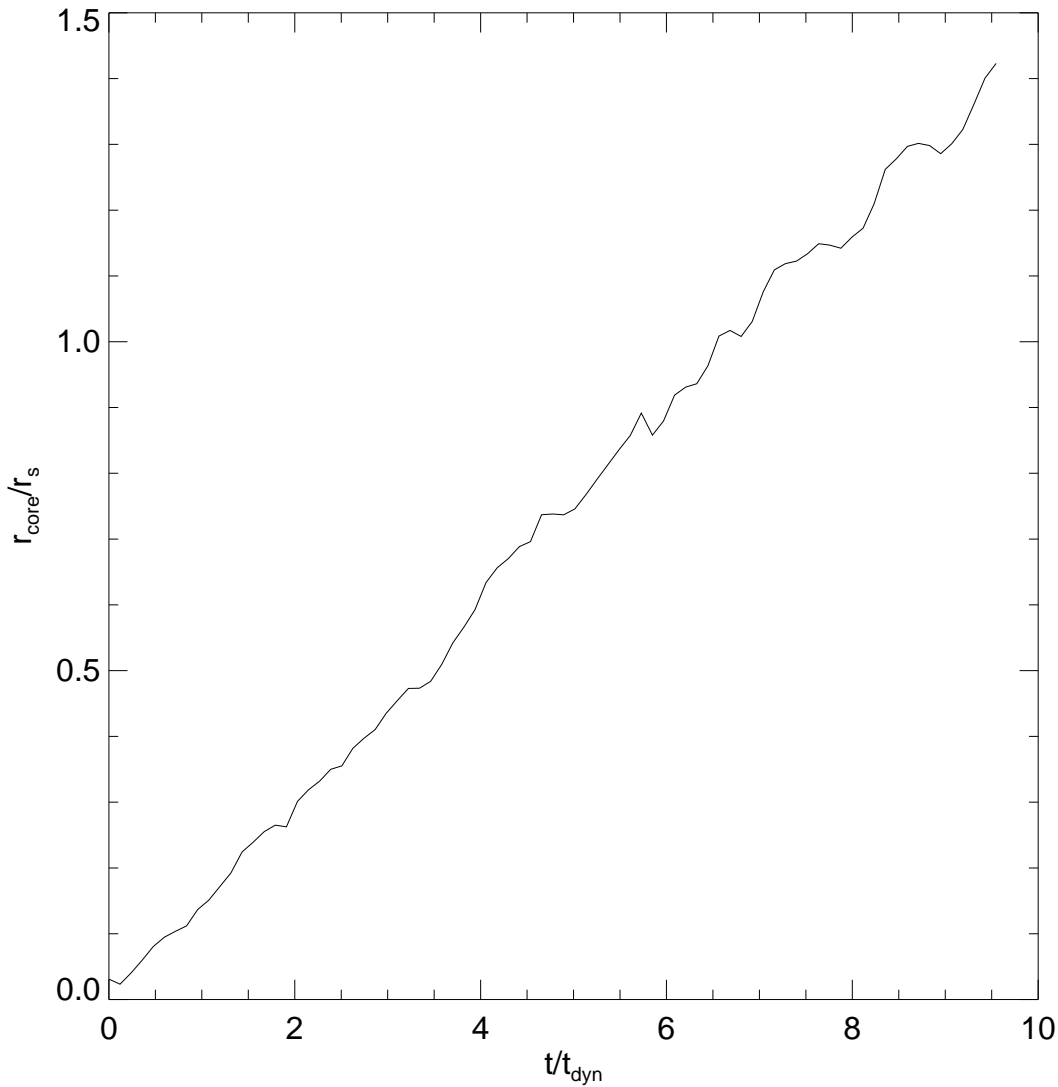}
  \caption{Growth of the core radius with time in the annihilating DM model}
  \label{fig:core-rad-growth}
\end{figure}

The halo shows a clear evolution from a Kaplinghat profile to a
modified Hernquist profile. At early times, the Kaplinghat model is a
better fit than the modified Hernquist model. At later times, the
modified Hernquist model is a better fit. This is good news, because
the modified Hernquist profile is closer in form to the Burkert
profile. The evolution in profile shape makes sense; early in the
simulation the adiabatic effect of annihilation on the gravitational
potential will be small, and the Kaplinghat profile should be a good
fit.  The core radius in the modified Hernquist profile continues to grow
throughout the simulation, as shown in Figure
\ref{fig:core-rad-growth}. The final core radius is 3~kpc, in the
range suggested by \citet{burkert:dwarfstruct:1995} for a halo of the
mass we studied.

\section{Discussion}
\label{sec:Conc}
Our simulations demonstrate that annihilation produce cores of radius
a few kiloparsecs in dark matter halos of dwarf galaxies; we now
discuss whether this dark matter model preserves the successes of
non-interacting CDM, the extent to which it addresses the problem of
too much substructure in galactic dark matter halos, and the
plausibility of the particle physics of the model.

\begin{figure}[htbp]
  \epsfig{file=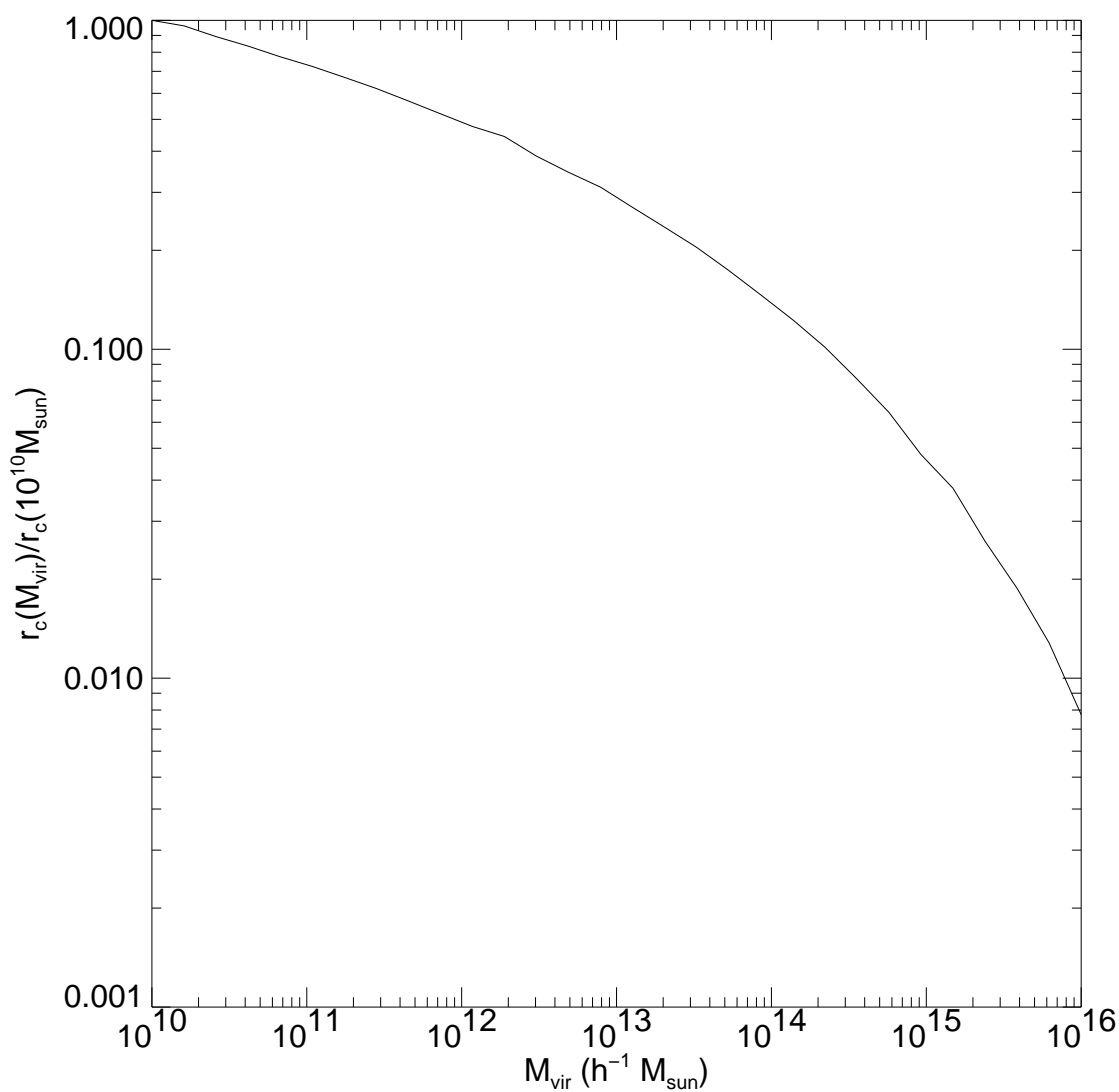} 
  \caption{Relative size of cores radius in halos of different mass.}
  \label{fig:core-mass-depend}
\end{figure}

There is little evidence for cores in clusters; for an annihilation
model to be plausible it must modify small scale behavior of dark
matter while leaving large scale behavior unchanged. In the analytic
model of \citet{kaplinghat:2000}, the core radius grows linearly in
time, and is given by
\begin{equation}
  \label{eq:kapl-core-rad}
  x_c = \frac{r_c}{r_s}= \rho_s \sigma_a t/m = \rho_s F\sigma^*_a t
\end{equation}
where $\rho_s$ is the characteristic density of the NFW profile and
$t$ is the age of the halo.  Since lower mass halos have higher
density, and formed earlier, than high mass halos, one would expect a
decreasing core radius with increasing mass. Figure
\ref{fig:core-mass-depend} shows the trend of core radius with mass,
assuming the core radius is given by equation
(\ref{eq:kapl-core-rad}); the core radius has been normalized to that
of a $10^{10} h^{-1}M_\odot$ halo (approximately 3~kpc) to eliminate
the dependence of the core radius on cross section.  Characteristic
densities and halo ages were calculated using the
\citet{eke:new-halo-den:2000} model, assuming a $\Omega_0=.3$,
$\Omega_{\Lambda}=1-\Omega_0 =.7$ flat cosmology with $h=.65$. Most of
the decrease in core radius is due to the decrease in characteristic
density as mass increases; the age difference of the lowest and
highest mass halo accounts for only 20\% of the difference in core
radius. Our model clearly creates large cores in low mass/high density
halos without producing them in high mass/low density halos.  Thus if
the annihilations lead to a core radius of 3~kpc for dwarf galaxies,
as appropriate, the predicted dark matter core radius in rich clusters
would be $\sim 120$~pc, which is deep within the central galaxy.

The core structure of the dark matter halo of a high surface
brightness galaxies is unclear.  \citet{eke:new-halo-den:2000} argue
that the concentrations of dark matter halos predicted in the standard
\lcdm\ model are consistent with both the Tully-Fisher relationship
and the amount of dark matter interior to the solar radius. On the
other hand, \citet{weiner:n4123b:2001} argues that detailed mass
modeling of the barred spiral NGC 4123 requires either a modest core
in the dark matter halo, with a core radius of 1--6~kpc, or an NFW
profile with very low concentration. Arguing more generally,
\citet{debattista:bars-core:2000,debattista:bar-core:1998} find that a
high density dark matter halo in a barred galaxy leads to rapid
slowing of bar rotation.  For the same choice of annihilation
cross-section that generates a 3~kpc core in dwarf galaxies, the core
radius in a $10^{12}$~M$_\odot$ galaxy such as the Milky Way would be
$\sim 1.2$~kpc, and would resolve the difficulties of bar spirals in
standard \lcdm\ discussed by \citet{debattista:bars-core:2000}.

The standard \lcdm\ model also predicts more substructure on galactic
scales than is observed
\citep{moore:substruct:1999,klypin:substruct:1999}. Although the
simulations we have done cannot directly address the question of
substructure, Figure \ref{fig:core-mass-depend} can be extended to low
mass halos. As halo mass shrinks, the core radius increases while the
virial radius increases. The two are equal for halos of mass
3$\times10^7$M$_{\odot}$; halos with mass $10^8$ solar masses have a
core radius that is half the virial radius. This suggests that halos
smaller than a few times $10^8$ solar masses would not be present in
an annihilating dark matter model. Further simulation of structure
formation in a cosmological context with annihilating
dark matter would be helpful.

One of the problems with earlier self-interacting dark matter proposals 
\citep{carlson:sidm:1992,machacek:perts:1994} is that ram pressure
stripping will destroy small halos too efficiently
\citep{delaix:nosidm:1995}. Newer models in which the dark matter
behaves like a viscosity-free fluid \citep{peebles:fluiddm:2000, peebles:fluiddm:1999,
  goodman:fluiddm:2000, riotto:quartic:2000} avoid this problem, but
may overly-suppress the growth of structure on small scales in the
linear regime \citep{peebles:fluiddm:2000}, a problem shared by the
earlier self-interacting dark matter models
\citep{machacek:perts:1994}.

The model proposed by \citet{carlson:sidm:1992} has some similarities
to our model. They investigated the properties of dark matter that
self-annihilated (that is, interactions occurred in which the
annihilation products were the dark matter particles). Their model
affects the structure of halos by heating dense regions, and affects
fluctuations in the early universe through the introduction of a
non-trivial Jean's length for the dark matter. Our model differs in
that we assume the dark matter annihilates to something relativistic
that no longer contributes to the gravitational potential of a halo
after annihilation. This should lead to a greater effect on the core
properties of the halo, through adiabatic changes to the gravitational
potential, than the model of \citet{carlson:sidm:1992}.
The interacting dark matter model proposed by \citet{spergel:sidm:2000}
is constructed so as to avoid excessive suppression of the power
spectrum on small scales. The dark matter is presumed to be cold, and
the interaction cross sections are small enough that the probability
of interaction is small at the time of recombination.

There are additional problems with an annihilating dark matter model.
First, the amount of energy released in dark matter annihilation is
tremendous; in our simulation of a single halo, the mass loss from
annihilation (in whatever the decay products are) has an energy
equivalent of $4.7\times 10^{45}$~erg~s$^{-1}$ averaged over the lifetime
of the universe.  This is far too large to be consistent with
observations of the gamma ray background.  If the
dark matter annihilates, the decay products cannot include photons.
Second, as discussed at length by \citet{kaplinghat:2000}, the
particle physics underlying an annihilating dark matter model must be
carefully contrived so as to avoid complete annihilation of the dark
matter in the early universe. The suppression of annihilation must be
in place for $z\ge 10000$, which is a rather low energy for any
reasonable phase transitions in the dark matter sector.

\section{Conclusion}

Annihilating dark matter leads to halo density profiles consistent
with those observed in dwarf spiral galaxies. Annihilation leads to a
rapid reduction in the central density of a dark matter halo, driven
mainly by adiabatic expansion of the core region as matter is
lost. An annihilation cross section sufficient to generate a core in a
dwarf galaxy halo generates a moderate core in a Milky Way sized halo,
and a negligible core in a rich cluster. 

The core radius decreases rapidly with halo mass, primarily because
larger objects form later in hierarchical clustering models, when the
mean density of the universe is lower. Objects smaller than $\sim
3\times 10^8$M$_\odot$ are unlikely to be present in an annihilating
dark matter model, because the core radius will be large compared to
the virial radius. A reduction in structure on this scale would be
useful; CDM predicts many more halos of this size than are observed.

On the other hand, there is no compelling motivation for such a
complex dark matter candidate. A special mechanism is needed to
suppress annihilation of the dark matter in the early universe, and
the dark matter must decay to particles that do not interact with
ordinary matter. Although a particle physics model with the required
properties could no doubt be constructed, such a model would be
contrived. 

Further study of the impact of annihilation in a cosmological setting
would be useful in evaluating the promise of this model.

\section{Acknowledgements}

{This work was supported in part by NSF grant AST00-71048.}

% End acknowledgements

%\bibliographystyle{elsart-harv}
%\bibliographystyle{apj}
%\bibliography{all}

%\begin{thebibliography}{7}
%%
%\bibitem{makino}
%\bibitem{burkert}
%\bibitem{moore}
%\bibitem{kochanek}
%\bibitem{turner}%

%\end{thebibliography}

\end{document}